\documentstyle[aps,prb,preprint]{revtex} 

\begin{document}  \preprint{\today} \draft
\title{Hole Photoproduction in Insulating
Copper Oxide}

\author{O. P. Sushkov\cite{Budker} }

\address{ School of Physics, The University of New South Wales,
    Sydney 2052, Australia}
\maketitle

\begin{abstract}
Basing on $t-J$ model we calculate the {\bf k}-dependence of a single hole
photoproduction  probability for CuO$_2$ plane at zero doping.
We also discuss  the radiation of spin-waves which can substantially
deform the shape of photoemission spectra.
\end{abstract}

\narrowtext 
Recent photoemission measurements\cite{Wells} for insulating Copper
Oxide Sr$_2$CuO$_2$Cl$_2$ give an unique possibility to determine single 
hole dispersion from experiment. However, to compare with data one 
needs to calculate the probability of the hole creation at a given 
momentum and energy transfer. The purpose of the present work is to 
calculate this probability in the framework of $t-J$ model and elucidate 
the related question with momentum dependence of quasihole residue (see 
Refs.\cite{Angelucci,Sorella}). In the conclusion we also 
comment on the shape of photoemission spectra.

It is widely accepted that low energy dynamics of CuO$_2$ planes in
Copper Oxides is described by $t-J$ model (see Ref.\cite{Dagotto} 
for a review).  At zero doping which corresponds to half filling this 
model is equivalent to Heisenberg antiferromagnet with long range 
N\'{e}el order in the ground state. We are interested in the process 
of an external photon kicking out an electron from the plane and creating 
a hole. The properties of a single hole are well established. For 
its description at $t/J \le 5$ one can use selfconsistent Born 
approximation\cite{Kane}. This approximation is justified by
the absence of a single loop correction to the spin-wave 
vertex\cite{Mart,Liu,Iga,Susf}. In some sence it is analog of the well 
known Migdal theorem for electron-phonon interaction. Hole dispersion 
has minima at ${\bf k}=(\pm \pi/2, \pm \pi/2)$ and it is almost 
degenerate along the face of magnetic Brillouin zone
$\gamma_{\bf k}= {1\over 2}(\cos k_x + \cos k_y) \ge 0$. The hole wave 
function $\psi_{{\bf k}\sigma}$ is characterized by quasimomentum 
${\bf k}$ defined inside the magnetic Brillouin zone, and pseudo-spin 
$\sigma =\uparrow \downarrow$. The pseudo-spin denotes the sublattice 
at which the hole is centered, and it is different from the usual spin. 
It is convenient to write the hole wave function in the form
\begin{equation}
\label{psi}
\psi_{{\bf k}\sigma}=h_{{\bf k}\sigma}^{\dag}|0 \rangle,
\end{equation}
where $|0 \rangle $ is the ground state of the Heisenberg model, and
$h_{{\bf k}\sigma}^{\dag}$ is the creation operator of the composite hole.
Let us also denote by $d_{n s}$ the annihilation operator of an electron 
at the cite $n$ and with usual spin $s=\uparrow \downarrow$. Let us 
fix the pseudo-spin $\sigma$ in Eq. (\ref{psi}), for example 
$\sigma =\uparrow$. 
The quasiparticle residue of the composite hole 
is given by 
\begin{equation}
\label{Z}
{\cal Z}_{\bf k} \propto \left|\langle \psi_{{\bf k}\uparrow}| 
\sum_{n \in up}
d_{n \uparrow} e^{i{\bf k}{\bf r}_n}|0\rangle \right|^2.
\end{equation}
We stress that the summation in this equation is restricted by one 
sublattice because the quasispin is fixed. This is the residue defined 
in the Refs.\cite{Kane,Mart,Liu,Iga,Susf}, and due to the Bloch theorem 
the exact relation 
\begin{equation}
\label{B}
{\cal Z}_{{\bf k} + {\bf Q}} = {\cal Z}_{\bf k}
\end{equation}
is valid for translation at the inverse vector of the magnetic 
sublattice, $({\bf Q}= (\pm \pi, \pm \pi))$.

When a photon kicks out an electron from the system it does not
separate the sublattices, therfore the external perturbation is  
\begin{equation}
\label{P}
\sqrt{2\over N} \sum_n
d_{n \uparrow} e^{i{\bf k}{\bf r}_n},
\end{equation}
and the probability of the process is 
proportional to
\begin{equation}
\label{A}
A_{\bf k} = \left|\langle \psi_{{\bf k}\uparrow}|\sqrt{2\over N} \sum_n
d_{n \uparrow} e^{i{\bf k}{\bf r}_n}|0\rangle \right|^2.
\end{equation}
Here $N$ is number of cites in the lattice. Normalization of the
external perturbation is rather arbitrary.  We choose it in such a 
way that $A_{\bf k}=1$ for Ising case and $t=0$. The value defined
by Eq.(\ref{A}) is different from the residue (\ref{Z}). 
For the calculation of 
$A_{\bf k}$ it is convenient to use $l=1$ string variational ansatz for 
the hole wave function\cite{String}
\begin{equation}
\label{h}
h_{{\bf k}\uparrow}^{(1)\dag}=\sqrt{2\over N}
\sum_{n \in \uparrow}\left(\nu_{\bf k}d_{n \uparrow} +
\mu_{\bf k}S_n^- \sum_{\bf \delta}\left[(1+v)-(u+v)\gamma_{\bf k}
e^{i{\bf k}{\bf \delta}}\right]d_{n+{\bf \delta}\downarrow}\right)
e^{i{\bf k}{\bf r}_n},
\end{equation}
where $\sum_{\bf \delta}$ denotes the summation over nearest neighbours,
and the coefficients are given by
\begin{eqnarray}
\label{numu}
\nu_{\bf k}&=&{1\over 2}\left[{{\Delta+2S_{\bf k}}\over{X S_{\bf k}}}
\right]^{1/2},
\nonumber\\
\mu_{\bf k}&=&{t\over{\left[Y S_{\bf k}(\Delta+2S_{\bf k})\right]^{1/2}}},\\
S_{\bf k}&=&\left(\Delta^2/4+4t^2\left[(1+y)-(x+y)\gamma_{\bf k}^2\right]
\right)^{1/2}.\nonumber
\end{eqnarray}
We set $J=1$. The parameters $\Delta =1.33$, $x=0.56$, $y=0.14$,
$X=0.8$, $Y=0.72$, $u=0.42$, $v=0.12$
are some combinations of the Heisenberg model spin-spin correlators
calculated in the Ref.\cite{String}. The residue (\ref{Z}) is 
proportional to the  coefficient $\nu_{\bf k}^2$ in the wave function 
(\ref{h}). We would like to stress that the 
string representation of the hole wave function is dual to the usual 
spin-wave representation, but the physical matrix elements are 
certainly independent of the representation  (see below). 

Keeping in mind the relations $d^{\dag}_{n\uparrow}d_{n\uparrow}=
{1\over 2}+S_z$ and $d^{\dag}_{n\downarrow}d_{n\uparrow}=S_n^-$
one can easily calculate the probability of photoemission (\ref{A})
in the minimal string approximation
\begin{equation}
\label{A1}
A^{(1)}_{\bf k}=\left[\nu_{\bf k}\left({1\over 2}+\sigma\right)
-8\mu_{\bf k}\left(1-u\right)q_1\gamma_{\bf k}\right]^2.
\end{equation}
Here $\sigma=\left| \langle 0|S_z|0\rangle\right|=0.3$ is average
magnetization, and 
$q_1={1\over 2}\langle 0|S_n^+S_{n+{\bf \delta}}^-|0\rangle = -0.08$
is the nearest neighbour spin-spin correlator for the Heisenberg model.
Due to the $\gamma_{\bf k}$ term in (\ref{A1})
$A^{(1)}_{{\bf k} + {\bf Q}} \ne A^{(1)}_{\bf k}$!
At zero hopping ($t=0$) $A^{(1)}_{\bf k}={1\over 2}+\sigma=0.8$.
It is smaller than 1 because of spin quantum fluctuations in the
initial state. This value agrees very well with result of spin-wave 
calculation\cite{Mal} and finite lattice computation\cite{Poil93}.

The $l=1$ string ansatz (\ref{h}) is a good approximation
for the hole wave function at $t \le 1$. For $t > 1$ further
dressing by $l > 1$ strings, or, in other words, multy
spin-wave virtual excitations is important. There are many
multy spin-wave components in the wave function of a
dressed hole. The  assumption  that these components do not give 
a coherent contribution in $A_{\bf k}$, but only reduce the probability 
of $h_{{\bf k}\sigma}^{(1)}$ configuration is quite reasonable. This is 
exactly what happens for the hole spin-wave vertex\cite{Susf}. Therefore
\begin{equation}
\label{AA}
A_{\bf k}= Z_{\bf k} A_{\bf k}^{(1)},
\end{equation}
where $Z_{\bf k}$ is the weight of $h_{{\bf k}\sigma}^{(1)}$
in the exact $h_{{\bf k}\sigma}$. The values of 
$Z_{{\bf k}=(\pi/2,\pi/2)} \approx Z_{\gamma_{\bf k}\approx 0}$ 
calculated in the Ref.\cite{Susf}
for different $t$ are presented in the Table I.
Note that the quasiparticle residue (\ref{Z}) equals to
${\cal Z}_{\bf k}=Z_{\bf k} \nu_{\bf k}^2 (1/2+\sigma)$,.
Using this equality one can also find $Z_{\bf k}$ from the results of 
Refs.\cite{Mart,Liu,Iga} where  quasiparticle residue ${\cal Z_{\bf k}}$
was calculated. In  Table I we present the ratio 
$A_{\bf Q}/A_0=A^{(1)}_{\bf Q}/A^{(1)}_0$ for different
values of $t$. We see that ${\bf k}$-dependence qualitatively
agrees with experimental data\cite{Wells}. We assume that
$t/J \ge 0$. It is interesting that at negative $t$ the ratio
$A_{\bf Q}/A_0$ is inverted. At $t \le 1$ our result for
$A_{\bf Q}/A_0$ reasonably agrees with that found by 
finite lattice computations\cite{Poil93,Poil931} and
numerical spin-wave analysis\cite{Angelucci}. However at
$t=3$ the ratio $A_{\bf Q}/A_0$ found in Refs.\cite{Poil93,Poil931}
for finite lattices is bigger than that from Table I.
Possible reason of this this disagreement is a variational
nature of $l=1$ string calculation. Therefore below we discuss
more regular way of calculation of $A_{\bf k}$. 

Let us use not string, but
spin-wave picture  considering hopping term by perturbation
theory\cite{Kane,Mart,Liu,Iga}. The wave function of a bare hole
(i.e. at $t=0$) is of the form 
$\psi^b_{{\bf k}\uparrow} \approx \sqrt{{2\over N}} 
\sum_{n \in \uparrow}
d_{n \uparrow} e^{i{\bf k}{\bf r}_n}|0\rangle$.
External perturbation is given by Eq. (\ref{P}), and we can
introduce bare vertex of single hole production. 
\begin{equation}
\label{Vbh}
V^b_h = \langle \psi^b_{{\bf k}\uparrow}|\sqrt{{2\over N}}
\sum_{n}
d_{n \uparrow} e^{i{\bf k}{\bf r}_n}|0\rangle \approx 1.
\end{equation}
Here we neglect the spin quantum fluctuations in the initial state.
External perturbation (\ref{P}) can also produce hole +
spin-wave final state. Let us denote by $\alpha_{\bf q}^{\dag}$
the creation operator of the spin-wave with momentum ${\bf q}$
and projection of spin $S_z=-1$ (see Ref.\cite{Manousakis}
for review). The vertex of hole + spin-wave creation equals
\begin{eqnarray}
\label{Vbhsw}
V^b_{h,sw} &=& \langle 0|\alpha_{\bf q} \left(\sqrt{{2\over N}}\sum_{m \in
\downarrow}d_{m \downarrow}^{\dag}
e^{-i({\bf k-q}){\bf r}_m} \right)
\left(\sqrt{{2\over N}}
\sum_{n}
d_{n \uparrow} e^{i{\bf k}{\bf r}_n}\right)|0\rangle =\nonumber\\
&=& {2\over N} 
\langle 0|\alpha_{\bf q} \sum_{m \in
\downarrow}S_m^-
e^{i{\bf q}{\bf r}_m}|0\rangle = \sqrt{{2\over N}}v_{\bf q},
\end{eqnarray}
where $v_{\bf q}$ is Bogoliubov parameter dioganalizing spin-wave
Hamiltonian: $u_{\bf q}=\sqrt{{1\over{\omega_{\bf q}}}+{1\over 2}}$, \
 $v_{\bf q}=-sign(\gamma_{\bf q})
\sqrt{{1\over{\omega_{\bf q}}}-{1\over 2}}$, \ 
 $\omega_{\bf q}=2\sqrt{1-\gamma_{\bf q}^2}$,
(see Ref.\cite{Manousakis})
We stress that (\ref{Vbhsw}) is a bare vertex. It corresponds to
instant production of hole + spin wave, but not production of hole
with subsiquent decay into hole + spin-wave. Note that $V^b_{h,sw}
\to \infty$ at ${\bf q} \to 0$. The reason is that perturbation
(\ref{P}) does not correspond any quasiparticle of the system,
and therfore usual Goldstone theorem is not applicable.
The verices  (\ref{Vbh}) and (\ref{Vbhsw}) are presented at Fig. 1.
Cross corresponds to external perturbation, solid line -
to the hole, and dashed  line - to the spin-wave.

Bare hole-spin wave vertex is equal to $g_{{\bf p},{\bf q}}=
\sqrt{{2\over N}}2f_0(\gamma_{\bf p}u_{\bf q}+\gamma_{\bf k}v_{\bf q})$,
with $f_0=2t$ (see e.g. Refs.\cite{Kane,Mart,Liu,Iga}). It corresponds to
the decay of the hole with momentum ${\bf k}$ into a hole
with momentum ${\bf p}={\bf k}-{\bf q}$ and a spin-wave with
momentum ${\bf q}$.

Now we can easily calculate the first correction to the production vertex
(\ref{Vbh}). It is given by the diagram presented at Fig.2.
\begin{eqnarray}
\label{Vh1}
&&\delta V_h^{(1)}(\epsilon =\epsilon_{\bf k},{\bf k})=
\sqrt{{2\over N}}\sum_{\bf q}{{v_{\bf q} g_{\bf k-q,q}}\over
{\epsilon_{\bf k}-\epsilon_{\bf k-q}-\omega_{\bf q}}}=\nonumber\\
&&
-{{8t}\over N}\sum_{\bf q}{{v_{\bf q} (\gamma_{\bf k-q} u_{\bf q}
+\gamma_{\bf k}v_{\bf q})}\over{\omega_{\bf q}}}=
{{4t}\over N}\gamma_{\bf k}\sum_{\bf q}\left({1\over{\omega_{\bf q}}}-
{1\over 2}\right)=0.4\cdot t \cdot \gamma_{\bf k}.
\end{eqnarray}
This gives
\begin{equation}
\label{Apt}
A_{\bf k}=\left(V_h^b+\delta V_h^{(1)}\right)^2=
(1+0.4 \ t \ \gamma_{\bf k})^2.
\end{equation}
It is in a very good agreement with string variational
result (\ref{A1}) which gives  $A_{\bf k}=
0.8(1+0.37 \ t \ \gamma_{\bf k})^2$  at $t \to 0$.

To perform spin-wave calculations at large $t$ one has to 
remember  that one loop correction to the hole-spin-wave
vertex is absent\cite{Mart,Liu,Iga,Susf} and therefore selfconsistent 
Born (SB) approximation\cite{Kane} is valid. It means that only
modification  in comparison with naive perturbation theory
is that in the diagram Fig. 2 one has to use dressed hole 
Green function found in selfconsistent  Born  approximation.
We carried out simplified
computation of this type with dressed hole Green function in
the form ${\cal Z}/(\epsilon -\epsilon_{\bf k}+i 0)$.
This is equivalent to the replacement of bare hole-spin-wave
coupling constant $f_0=2t$ with the effective one $f={\cal Z}\cdot f_0$,
see Ref.\cite{Susf}. The results of this computation at $t \le 4$
reasonably agree with variational answer (\ref{A1}), (\ref{AA}).
One can certainly calculate $A_{\bf k}$ more accurately using
hole Green function found in Refs.\cite{Mart,Liu,Iga}.
However, we would like to stress that for detailed analysis
of photoemission spectra $A_{\bf k}$ is not enough.

As we already discussed an external photon can produce
hole + spin-wave as well as a single hole. Therefore the
$\epsilon, {\bf k}$ dependence of photoemission spectrum
is given by the imaginary part of diagrams presented
at Fig.3, with solid line be hole Green function 
in selfconsistent  Born  approximation $G_{SB}$. The
Fig. 3 represents new Green function corresponding to the
external perturbation (\ref{P}).
The diagram Fig.3a gives usual $Im \ G_{SB}(\epsilon,{\bf k})$. The
diagram Fig. 3b takes into account already discussed
difference of $A_{\bf k}$ from pure residue ${\cal Z}_{\bf k}$,
but still this contribution is proportional to $Im \ G_{SB}$.
The diagram Fig.3c gives nonresonant background caused by 
the emission of spin-waves. And finally the diagrams Fig3d,e
represent interference between background and resonant
production. This interference shifts maximum of photoemissin 
spectrum from $\epsilon = \epsilon_{\bf k}$. Whether this shift
could explain the experimental spectra\cite{Wells} is still an 
open question.

Before conclusion we would like to note that the ``electron''
creation operator is actually Zhang and Rice singlet creation
operator.\cite{Zhang}. Therfore formfactor of this singlet
$F_{\bf k}$ should be also taken into account. Calculation of 
$F_{\bf k}$ is not a problem because the wave function of the
singlet is now well known\cite{Bel}.

\vskip2ex
In the present work we calculate single hole photoproduction
probability $A_{\bf k}$ in frameworks of $t-J$ model. In 
agreement with experiment $A_{\bf k}$ drops down outside
magnetic Brillouin zone. We also point out that shape of
photoemission spectrum is different from $Im \ G_{SB}(\epsilon,{\bf k})$,
and due to the interference the difference is not just a smooth
background.

\vskip2ex
I am very grateful to D. Poilblanc and G. Sawatzky for
attracting my attention to this problem.
This  work has been
done during my stay at the Laboratoire de Physique Quantique,
Universite Paul Sabatier. I am gratefully acknowledge the
hospitality and financial support.

\tighten

\begin{table}
\caption{The weight $Z_{\gamma_{\bf k}\approx 0}$
of $l=1$ string configuration in the dressed hole wave function,
and
the ratio of a single hole creation probabilities $A_{\bf k}$
at  ${\bf k}={\bf Q}=(\pm \pi,\pm \pi)$ and ${\bf k}=0$
for different values of $t/J$}

\begin{tabular}{l l l l l l l l}
\multicolumn{1}{c}{$t/J$} &
\multicolumn{1}{c}{0} &
\multicolumn{1}{c}{0.2} &
\multicolumn{1}{c}{0.5} &
\multicolumn{1}{c}{1} &
\multicolumn{1}{c}{2} &
\multicolumn{1}{c}{3} &
\multicolumn{1}{c}{4} \\ \hline

$Z$&1& 0.96 & 0.88 & 0.79 & 0.63 & 0.53&0.45 \\
$A_{\bf Q}/A_0$&1& 1.3 & 1.9 & 2.5 & 3.3 & 3.7 & 3.9 \\
\end{tabular} \end{table}

\vskip10ex

Figure captions,\\
Fig.1 Bare vertices of a single hole creation and hole + spin-wave creation.
The cross corresponds to external perturbation, solid line - to the hole,
and dashed line - to the spin-wave.\\
Fig.2 First order correction to the single hole creation amplitude.\\
Fig.3 The diagrams contributing into photoemission spectrum. Solid line
corresponds to the hole Green function in selfconsistent Born approximation..

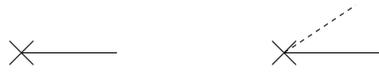
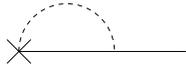
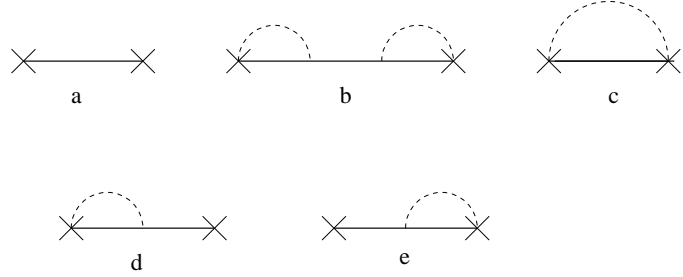
\end{document}